\documentclass[]{interact}

\usepackage{epstopdf}
\usepackage[caption=false]{subfig}

\usepackage[numbers,sort&compress]{natbib}
\usepackage{amsmath,amssymb,amsfonts}
\usepackage{mathrsfs}
\usepackage{chemformula}
\usepackage{color}
\usepackage[normalem]{ulem}
\usepackage{stmaryrd}

\usepackage[parfill]{parskip} 
\graphicspath{{./figures/}}

\bibpunct[, ]{[}{]}{,}{n}{,}{,}
\renewcommand\bibfont{\fontsize{10}{12}\selectfont}

\theoremstyle{plain}

\theoremstyle{definition}

\theoremstyle{remark}

\newcommand{\blue}[1]{\textcolor{black}{#1}}

\newcommand{\beq}{\begin{equation}}
\newcommand{\eeq}{\end{equation}}
\newcommand{\pfr}[2]{\ensuremath{\frac{\partial #1}{\partial #2}}}
\newcommand{\pfi}[2]{\ensuremath{{\partial #1}/{\partial #2}}}

\newcommand\Pec{\mbox{\textit{Pe}}}

\newcommand\Lew{\mbox{\textit{Le}}}

\newcommand{\vect}[1]{\mathbf{#1}}

\begin{document}


\title{Diffusive-thermal instabilities of a planar premixed flame aligned with a shear flow}

\author{
\name{Joel Daou$^*$\thanks{Corresponding author. Email: joel.daou@manchester.ac.uk} and Prabakaran Rajamanickam}
\affil{Department of Mathematics, University of Manchester, Manchester M13 9PL, UK}
}

\maketitle

\begin{abstract}
The stability of a thick planar premixed flame, propagating steadily in a direction transverse to that of unidirectional shear flow, is studied. A linear stability analysis is carried out in the asymptotic limit of infinitely large activation energy, yielding a dispersion relation. The relation characterises the coupling between Taylor dispersion (or shear-enhanced diffusion) and the flame thermo-diffusive instabilities, in terms of two main parameters, namely, the reactant Lewis number $\Lew$ and the flow Peclet number $\Pec$. The implications of the dispersion relation are discussed and various flame instabilities are identified and classified in the $\Lew$-$\Pec$ plane. An important original finding is the demonstration that for values of the Peclet number exceeding a critical value, the classical cellular instability, commonly found for $\Lew<1$, exists now for $\Lew>1$ but is absent when $\Lew<1$.  In fact, the cellular instability identified for $\Lew>1$ is shown to occur either through a finite-wavelength stationary bifurcation (also known as type-I$_s$) or through a longwave stationary bifurcation (also known as type-II$_s$). The latter type-II$_s$ bifurcation leads in the weakly nonlinear regime to a Kuramoto-Sivashinsky equation, which is determined. As for the oscillatory instability, usually encountered in the absence of Taylor dispersion in $\Lew>1$ mixtures, it is found to be absent if the Peclet number is large enough. The stability findings, which follow from the dispersion relation derived analytically, are complemented and examined numerically for a finite value of the Zeldovich number. The numerical study involves both  computations of the eigenvalues of a linear stability boundary-value problem and numerical simulations of the time-dependent governing partial differential equations. The computations are found to be in good qualitative agreement with the analytical predictions.
\end{abstract}


\section{Introduction}
\label{sec:intro}

Thick premixed flames propagating in shear flows, such as flames propagating in a narrow-channel Poiseuille flow, are subject to enhanced diffusion in the longitudinal flow direction due to Taylor's dispersion mechanism, as reported in recent investigations~\cite{pearce2014taylor,daou2018taylor,daou2021effect,rajamanickam2022thick}. Meanwhile, diffusion transport in a direction transverse to the flow direction is  only associated with molecular diffusion. Therefore, diffusion appears as effectively anisotropic which in turn significantly affect flame propagation and  stability. The thermo-diffusive stability of premixed flames in the presence of Taylor dispersion has been recently studied, analytically and numerically, in the case of flames propagating in the longitudinal (or flow) directions. This has been done assuming adiabatic conditions in~\cite{daou2021effect} and accounting for heat losses in~\cite{daou2023flame}. The equivalent stability problem for flames aligned with the shear flow, propagating in a direction transverse to the flow has not been addressed yet. This problem is important to investigate on account of the anisotropy of diffusion aforementioned, which is expected to markedly alter the previous  findings. Another motivation to consider the problem of transverse premixed-flame propagation is due to the interesting findings of a recent investigation on the effect of Taylor dispersion on nonpremixed flames aligned with the direction of a shear flow~\cite{rajamanickam2023stability}. This investigation demonstrates that Taylor dispersion can lead unexpectedly to cellular instability of a planar diffusion flame in mixtures with Lewis numbers above unity. This result has been argued in~\cite{rajamanickam2023stability} to present a plausible mechanism to explain the formation of so-called \textit{diffusion flame streets} observed experimentally in non-premixed microcombustors~\cite{miesse2005diffusion,miesse2005experimental,xu2009studies}. The question arises naturally therefore whether a similar cellular instability can occur for large Lewis numbers in the premixed case; namely, for flames propagating in the transverse direction, noting that such cellular instability for large Lewis numbers does not occur for flames propagating in the longitudinal direction~\cite{daou2021effect,daou2023flame}. One of the main objectives of this paper is to provide a clear answer to this question; as well as to describe the instabilities encountered and identify the conditions for their occurrence in terms of the parameters, mainly the Lewis number characterising the mixture and the Peclet number characterising the flow. In order to focus the analysis on these objectives, mainly concerned with the coupling between Taylor dispersion and the thermo-diffusive flame instabilities~\cite{sivashinsky1977nonlinear}-\cite[pp.~477-479]{clavin2016combustion}, we shall adopt the thermo-diffusive approximation of constant density and constant transport properties and neglect further the effect of heat losses.

The paper is structured as follows. The problem formulation is given in $\S$\ref{sec:form} within a Hele-Shaw or two-dimensional channel configuration\cite{joulin1994influence,fernandez2018analysis,al2019darrieus}, involving a unidirectional shear flow. In this configuration, the depth-averaged governing equations are written down and account for anisotropic diffusion, with enhanced diffusion in the longitudinal flow direction. A reformulation of the problem suitable for the limit of infinite Zeldovich number, $\beta\rightarrow \infty$,  is provided  in $\S$\ref{sec:nef}. This reformulation is used in the linear stability analysis of $\S$\ref{sec:linear}, which is carried out analytically and culminates in the derivation of a dispersion relation. The implications of the dispersion relation are discussed in detail in $\S$\ref{sec:regime}, where various flame instabilities are identified and classified in the parameters space. The stability findings, which follow from the dispersion relation derived analytically in the limit $\beta\rightarrow \infty$, are then complemented and examined numerically in $\S$\ref{sec:numerical} for a finite value of the Zeldovich number, $\beta=10$. The numerical study in $\S$\ref{sec:numerical} involves both the computation of the eigenvalues of a linear stability boundary-value problem and the numerical simulations of the time-dependent governing partial differential equations for illustrative cases. Concluding remarks are provided in $\S$\ref{sec:conclusions}.

\section{Problem formulation}
\label{sec:form}

\begin{figure}[h!]
\centering
\includegraphics[scale=0.8]{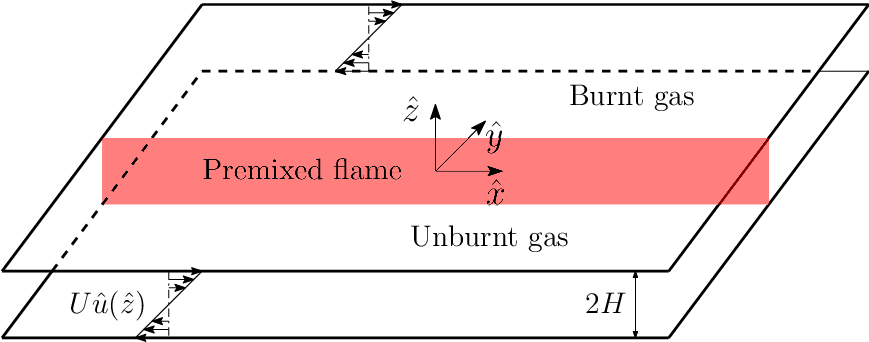}
\caption{Planar premixed flame in a channel of width $2H$, aligned with the direction ($\hat x$) of a shear flow  and propagating in the negative transverse direction ($\hat y$).} 
\label{fig:Setup}
\end{figure}

Consider a narrow channel as depicted in Fig.~\ref{fig:Setup} in which a planar premixed flame is propagating in the negative $\hat y$-direction with laminar flame speed $S_L$, in the presence of a shear flow. The shear flow is assumed, for simplicity, to have zero mean such as in the case of a Couette flow. Alternatively, in the case of Poiseuille flow for example, the frame of reference is chosen to be moving with the mean flow speed in the $\hat x$-direction. The frame is furthermore assumed to move with speed $S_L$ in the negative $\hat y$-direction, so that the flame may be considered steady in the unperturbed state. In other words, the flow field adopted is of the form
\begin{equation}
    \hat v_x= U \hat u (\hat z), \qquad \hat v_y=S_L, \qquad \hat v_z=0,
\end{equation}
where $U$ is the flow (maximum) amplitude and $\hat u(\hat z)$ is the scaled zero-mean shear-flow profile, such that e.g. $\hat u= \hat z /H$ for a Couette flow and $\hat u = 1/3-\hat z^2/H^2$ for a plane Poiseuille flow.

To model the chemistry, a single-step irreversible Arrhenius reaction is adopted, with pre-exponential factor $B$, activation energy $E$ and heat release $q$ per unit mass of fuel consumed. For sufficiently fuel-lean conditions, the fuel burning rate per unit volume can be written as $\rho B Y_F e^{-E/RT}$, where $\rho$ is the density assumed to be constant, $R$ is the universal gas constant, $Y_F$ is the fuel mass fraction and $T$ is the temperature. The adiabatic flame temperature $T_{ad}$, the Zeldovich number $\beta$ and the heat release parameter $\alpha$ are defined by
\begin{equation}
    T_{ad} = T_u + \frac{qY_{F,u}}{c_p}, \qquad \alpha = \frac{T_{ad}-T_u}{T_{ad}}, \qquad \beta = \frac{E(T_{ad}-T_u) }{RT_{ad}^2}    \nonumber
\end{equation}
where $T_u$ and $Y_{F,u}$ are the temperature and fuel mass fraction in the unburnt gas and $c_p$ is the specific heat at constant pressure.

For non-dimensionalization, we shall use the laminar flame thickness $\delta_L=D_T/S_L^0$ as unit length and $\delta_L/S_L^0$ as unit time, where $D_T$ is the thermal diffusivity (assumed constant). Here $S_L^0$ refers to the laminar flame speed (for $\beta\rightarrow \infty$) which is given by $S_L^0= [2\beta^{-2} \Lew BD_T \exp(-E/RT_{ad})]^{1/2}$, where $\Lew$ is the (fuel) Lewis number. The dependent variables $Y_F$ and $T$ are normalized by introducing
\begin{equation}
    y_F = \frac{Y_F}{Y_{F,u}}, \qquad \theta = \frac{T-T_u}{T_{ad}-T_u}.
\end{equation}

\newpage
 We consider the limit $H\ll \delta_L$ which leads upon averaging in the $\hat z$-direction to a problem which is effectively two-dimensional in the first approximation, as shown in~\cite{pearce2014taylor,daou2018taylor,daou2021effect,rajamanickam2022thick}. The non-dimensional depth-averaged equations read
\begin{align}
    \pfr{y_F}{t} + S \pfr{y_F}{y}&= \frac{1}{\Lew}\left[(1+p^2\Lew^2) \pfr{^2y_F}{x^2} + \pfr{^2y_F}{y^2}\right] - \omega, \label{yF}\\
    \pfr{\theta}{t} + S \pfr{\theta}{y}&= (1+p^2) \pfr{^2\theta}{x^2} + \pfr{^2\theta}{y^2} + \omega \label{theta}
\end{align}
where $p=\sqrt{\gamma}\Pec$ is a parameter proportional to the Peclet number $\Pec$,
\begin{equation}
    S = \frac{S_L}{S_L^0}, \qquad \Pec = \frac{UH}{D_T}, \qquad \omega = \frac{\beta^2}{2\Lew}y_F\exp\left[\frac{\beta(\theta-1)}{1+\alpha(\theta-1)}\right]. \nonumber
\end{equation}
The constant $\gamma$ is a numerical coefficient, which is determined by the shear-flow profile and is given by
\begin{equation}
\gamma = \int_0^1dz'\left[\int_0^{z'}dz\,\hat u(z)\right]^2,   \nonumber  
\end{equation}
where $z=\hat z/H$, so that $\gamma=8/945$ for the Poiseuille flow and $\gamma=1/20$ for the Couette flow, introduced above. We note that the parameter $p$ quantifies the enhancement of diffusion by Taylor dispersion and that this enhancement is in the longitudinal $x$-direction, but not in the transverse $y$-direction. The boundary conditions in the $y$-direction are given by
\begin{align}
    y_F=1, \quad \theta=0 &\quad \text{as} \quad y\rightarrow -\infty, \label{BCunburnt}\\
    y_F=0, \quad \pfr{\theta}{y}=0 &\quad \text{as} \quad y\rightarrow +\infty. \label{BCburnt}
\end{align}
The main focus of this investigation is the determination of steady, one-dimensional solutions of the problem~\eqref{yF}-\eqref{BCburnt} independent of $x$ and $t$ and their stability. The stability analysis is addressed analytically in the asymptotic limit of infinitely large Zeldovich number in $\S$\ref{sec:nef}-$\S$\ref{sec:regime} and the predictions are then examined numerically for finite value of $\beta$ in $\S$\ref{sec:numerical}.

\section{Formulation in the near-equidiffusional flame (NEF) limit}
\label{sec:nef}

The stability analysis will be carried out asymptotically in the limit $\beta\rightarrow \infty$ using the so-called near-equidiffusional flame (NEF) approximation based on the assumption that the Lewis number deviates little from unity \cite[p.~33]{buckmaster1983lectures}. Within this approximation, the reduced Lewis number $l \equiv \beta \left(\Lew-1\right)$ is $O(1)$ as $\beta\rightarrow \infty$ and equations~\eqref{yF}-\eqref{theta} can be in terms of the leading-order temperature $\theta^0$ and $h\sim\beta(y_F+\theta^0-1)$ as
\begin{align}
    \theta^0_t + S \theta^0_y &= (1+p^2) \theta^0_{xx}+ \theta^0_{yy}, \label{thetanef}  \\
    h_t + Sh_y &= (1+p^2) h_{xx}+ h_{yy} + l[(1-p^2)\theta^0_{xx} + \theta^0_{yy}] \label{hnef}
\end{align}
which are applicable outside an infinitely thin reaction sheet, given by $y=f(x,t)$ say. The equations are subject to the boundary conditions
\begin{align}
    \theta^0=0, \quad h=0 &\quad \text{as} \quad y\rightarrow -\infty\\
    \theta^0=1, \quad h\, \text{is finite} &\quad \text{as} \quad y\rightarrow +\infty
\end{align}
and the jump conditions
\begin{subequations}  \label{jumpCons}
\begin{eqnarray}
 && \left \llbracket\theta^0 \right \rrbracket =0\,, \, \quad \left \llbracket h \right \rrbracket=0  \label{e:JC1}\\
 && \left \llbracket h_y \right \rrbracket+ \frac{1+f_x^2(1-p^2) }{1+f_x^2(1+p^2)} \, l \, \left \llbracket \theta^0_y \right \rrbracket =0
  \label{e:JC2}\\
 && \sqrt{1+f_x^2(1+p^2)} \,  \left \llbracket \theta^0_y \right \rrbracket = - e^{h/2}
  \label{e:JC3}
\end{eqnarray}
\end{subequations}
applicable at $y=f(x,t)$. Here we have used  the notation $\left \llbracket \,\varphi\, \right \rrbracket= \varphi |_{y=f^+} - \varphi|_{y=f^-}$. It is worth pointing out that jump conditions~\eqref{jumpCons}, which account for the presence of Taylor dispersion, are derived from an analysis of the structure of the reaction zone. Since the derivation is similar to that reported in~\cite{daou2018taylor}, its details are not included here.

\section{Linear stability analysis for $\beta\rightarrow \infty$}
\label{sec:linear}

We examine herein the stability of the steady planar flame solution, denoted by an overbar, which satisfies equations (\ref{thetanef})--(\ref{jumpCons}) with $\partial/\partial t=0$ and $\partial/\partial x=0$ and is given by
\begin{equation} 
    S = 1, \qquad \bar{f}=0,\qquad  \bar\theta = \begin{cases}
    e^{y} &\text{for}\,\, y<0 \\
    1 &\text{for}\,\, y>0
    \end{cases},
    \qquad \bar h = \begin{cases}
    -ly e^{y} &\text{for}\,\, y<0 \\
    0 &\text{for}\,\, y>0
    \end{cases}.
\end{equation}
To this solution, we add infinitesimal normal-mode disturbances such that
\begin{align} \label{perturbation}
          \begin{bmatrix}
           f \\
           \theta^0\\
           h
          \end{bmatrix} =
          \begin{bmatrix}
           0 \\
           \bar{\theta}(y) \\
           \bar{h}(y)
         \end{bmatrix} + \,  e^{\sigma t +i k x}
         \begin{bmatrix}
           \tilde f \\
           \tilde{\theta}(y)  \\
          \tilde{h}(y)
          \end{bmatrix}
  \end{align}
where $k$ denotes the real wavenumber and $\sigma$ a constant (not necessarily real) characterizing the growth rate/frequency of the perturbation. 

The methodology used to obtain the dispersion relation $\varphi(\sigma,k,l,\lambda)=0$ is classical, see e.g.~\cite{sivashinsky1977nonlinear}. We begin by substituting the perturbed solutions~\eqref{perturbation} into equations~\eqref{thetanef}--\eqref{jumpCons}. This leads to an eigenvalue boundary problem for the functions $\tilde{\theta}(y)$ and $\tilde{h}(y)$  which is given by
\begin{align}
    \tilde \theta_{yy} -  \tilde \theta_y - [\sigma + k^2(1+p^2)]\tilde\theta &=0,\\
     \tilde h_{yy} -  \tilde h_y - [\sigma + k^2(1+p^2)]\tilde h &= -l [\tilde \theta_{yy}+k^2(p^2-1) \tilde \theta]
\end{align}
applicable for $y\neq 0$. These equations are subject to the boundary conditions
\begin{align}
    \tilde\theta=0, \quad \tilde h = 0 \quad \text{as} \quad y\rightarrow \pm \infty,    
\end{align}
and the linearised jump conditions
\begin{align}
       \llbracket\tilde\theta \rrbracket=\tilde f, \quad  \llbracket\tilde h \rrbracket = -l\tilde f, \quad  \llbracket\tilde h_y \rrbracket + l  \llbracket\tilde\theta_y \rrbracket = - l\tilde f, &\quad  \llbracket\tilde\theta_y \rrbracket = \tilde f - \frac{1}{2}\tilde h(0^+) \label{jumpappendix}
\end{align}
applicable at $y=0$, i.e., $\left \llbracket \,\varphi\, \right \rrbracket= \varphi |_{y=0^+} - \varphi|_{y=0^-}$.

The solution for $\mathrm{Re}\{\sigma+k^2(1+p^2)\}>0$ is given by 
\begin{equation}
    \tilde \theta = -\tilde f\begin{cases}
         e^{(1+\Gamma)y/2}  \\
        0 
    \end{cases}, \quad
    \tilde h = \tilde f\begin{cases}
        [1-\Gamma + l(1+\chi y)] e^{(1+\Gamma)y/2} & \text{for}\,y<0 \\
        (1-\Gamma) e^{(1-\Gamma)y/2} & \text{for}\,y<0
    \end{cases}
\end{equation}
with $\chi = (1+\Gamma)^2/4\Gamma + \kappa^2(2\lambda-1)/\Gamma$, satisfying the solvability condition or the dispersion relation 
\begin{equation}
    2\Gamma^2(1-\Gamma) + l(1-\Gamma+2\sigma+4\lambda \kappa^2)=0. \label{disp}
\end{equation}
Here
\begin{equation}
    \Gamma = \sqrt{1+4\sigma+4\kappa^2}, \quad \kappa = k\sqrt{1+p^2}, \quad \lambda = \frac{p^2}{1+p^2} \in [0,1]. \label{gamma}
\end{equation}

Equation \eqref{disp} can be solved for $\sigma=\sigma(\kappa;l,\lambda)$, yielding in general three roots $\sigma_1$, $\sigma_2$ and $\sigma_3$, or less. Clearly, the stability of a mode with wavenumber $\kappa$ is dictated by the root $\sigma$ whose real part is equal to $\max(\mathrm{Re}\{\sigma_1\},\mathrm{Re}\{\sigma_2\},\mathrm{Re}\{\sigma_3\})$. We shall denote this eigenvalue by
\begin{equation}
   \sigma_{\mathrm{max}}= \sigma_{\mathrm{max}}(\kappa;l,\lambda). 
\end{equation}
Furthermore, $\mathrm{Re}\{\sigma_{\mathrm{max}}\}$ attains a maximum value at $\kappa=\kappa_m$ say, with corresponding complex growth rate $\sigma=\sigma_m$. These values characterise the most unstable mode and are functions of $l$ and $\lambda$,
\begin{equation}
  \sigma_m=\sigma_m(l,\lambda)  \qquad \text{and} \qquad \kappa_m=\kappa_m(l,\lambda). \label{sigmam}
\end{equation}

\section{Implications of the dispersion relation}
\label{sec:regime}

In this section, we will examine the implications of the dispersion relation~\eqref{disp} on flame stability using the notations~\eqref{gamma}-\eqref{sigmam}. 

\begin{figure}[h!]
\centering
\includegraphics[scale=0.6]{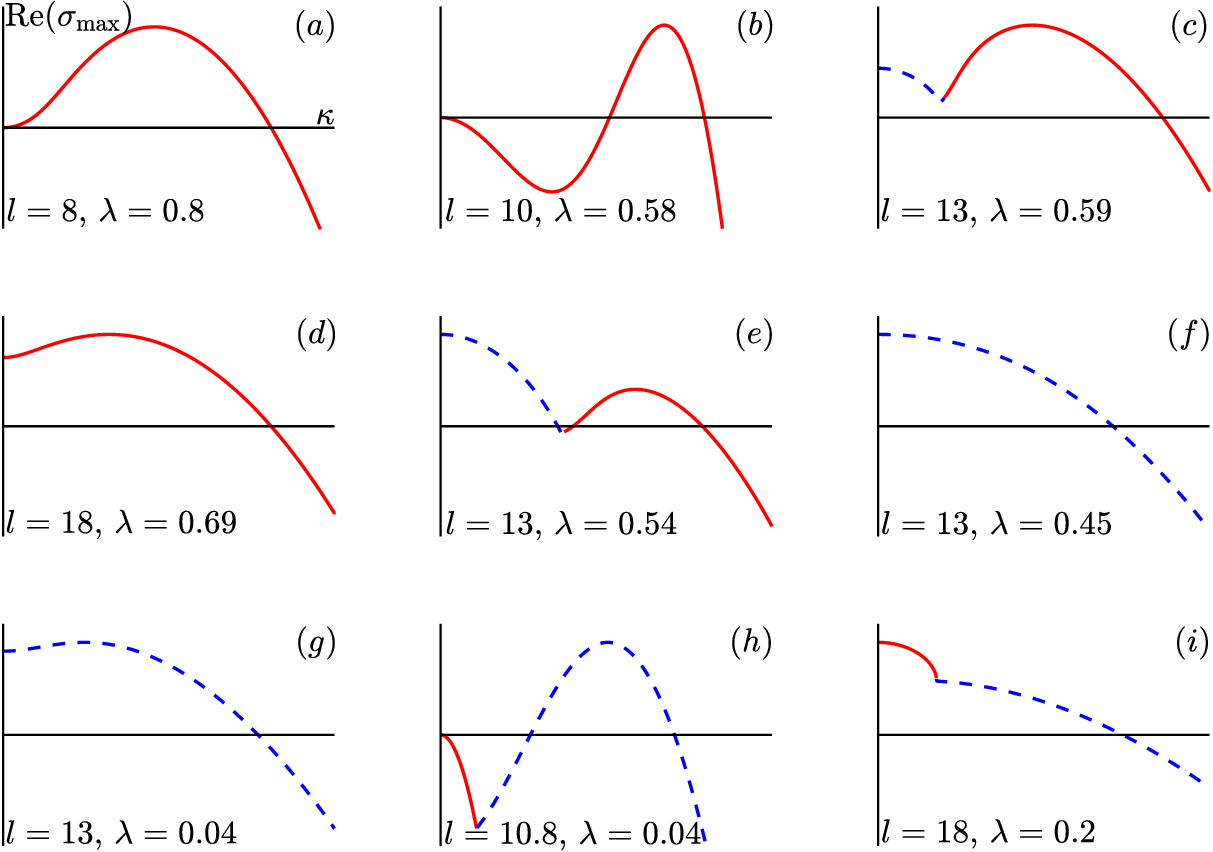}
\caption{Growth rate $\mathrm{Re}(\sigma_{\mathrm{max}})$ versus wavenumber $\kappa$ for selected values of $l$ and $\lambda$. Solid lines indicate that $\sigma_{\mathrm{max}}$ is real ($\mathrm{Im}(\sigma_{\mathrm{max}})=0$) and dashed lines indicate that $\sigma_{\mathrm{max}}$ has a non-zero imaginary part ($\mathrm{Im}(\sigma_{\mathrm{max}})\neq 0$). The horizontal and vertical scales are not indicated and are chosen individually for each subfigure for the sake of clarity.} \label{fig:nine}
\end{figure}

\subsection{Illustrative cases and terminology}

The function $ \sigma_{\mathrm{max}}(\kappa)$ assumes various forms depending on the value of $l$ and $\lambda$, each one resembling one of the dispersion curves depicted in Fig.~\ref{fig:nine}. In this figure, a first category of dispersion curves can be identified for which the maximum growth rate $\sigma_m$ is real and occurs for a non-zero value of the wavenumber, $\kappa_m\neq 0$; this is the case of subfigures~\ref{fig:nine}$(a)$ to \ref{fig:nine}$(d)$. In this category, the instability may be termed as a \textit{cellular instability} since the unstable planar flame solutions are expected to evolve into cellular structures, at least near the instability onset. A second category of dispersion curves can be identified for which the instability may be characterised as being an \textit{oscillatory instability}, since it corresponds to $\sigma_m$ having a non-zero imaginary part. This is the case of subfigures~\ref{fig:nine}$(e)$ to \ref{fig:nine}$(h)$. The oscillatory instability appears here as a longwave instability when $\kappa_m=0$ as in subfigures~\ref{fig:nine}$(e)$ and~\ref{fig:nine}$(f)$ or as a finite-wavelength instability when $\kappa_m\neq 0$ as in subfigures~\ref{fig:nine}$(g)$ and~\ref{fig:nine}$(h)$. Note that in the case of subfigure~\ref{fig:nine}$(e)$, a finite wavelength cellular instability can occur instead of the oscillatory longwave instability, if the domain is not large enough. There are also cases such as in subfigure~\ref{fig:nine}$(i)$ where $\kappa_m=0$ and $\sigma_m$ is real, while $\sigma_{\mathrm{max}}(\kappa)$ has a non-zero imaginary part except close to $\kappa=0$. Solutions for such cases exhibit a non-oscillatory longwave instability near onset, at least in sufficiently large domains; in smaller domains, oscillations corresponding to non-zero wavenumbers are however expected near onset.


\subsection{Stability regime diagram}

\begin{figure}[h!]
\centering
\advance\leftskip-0.8cm
\advance\rightskip-0.9cm
\includegraphics[scale=0.66]{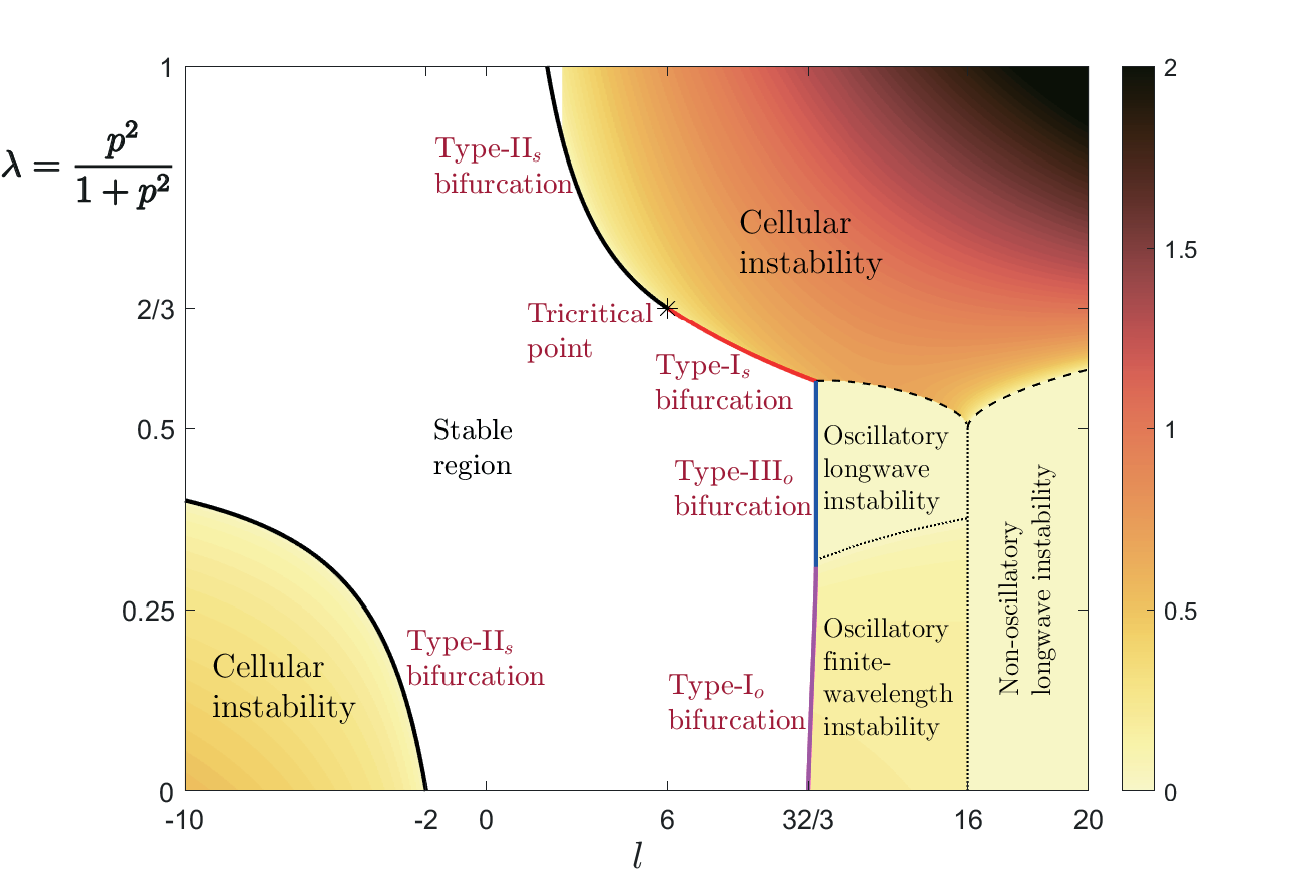}
\caption{Stability regime diagram of a premixed flame propagating transversely to a shear flow in the $l$-$\lambda$ plane, which is equivalent to the $\Lew$-$\Pec$ plane. The figure is based on the dispersion relation~\eqref{disp}, obtained in the limit $\beta\to\infty$. The colour scale  represents the wavenumber $\kappa_m$ of the most unstable mode (with complex growth rate $\sigma_m$). The types of the bifurcation curves, represented by solid lines, are discussed in $\S$\ref{sec:bifur}.} 
\label{fig:regime}
\end{figure}

A convenient way of summarizing the various stability results is to delimit instability regions in an $l$-$\lambda$ plane, which is equivalent to $\Lew$-$\Pec$ plane, as done in Fig.~\ref{fig:regime}. The colour scale in this figure represents the wavenumber $\kappa_m$ of the most unstable mode (with complex growth rate $\sigma_m$). The solid lines in this figure separate the stable (white) region from the unstable ones, whereas the dashed lines\footnote{The dashed curves meet at a cusp located at $(l,\lambda)=\left(16,\frac{1}{2}\right)$. The crossing from the non-cellular to the cellular region as $\lambda$ is increased involves a discontinuous jump in $\kappa_m$ for $l<16$ and a continuous transition for $l>16$. Similar transitions have been observed in diffusion flames which have been described with more detail in~\cite{rajamanickam2023stability}.}  for $l>0$ separate the region of cellular instability (above these lines) from that of non-cellular instability. Note that the region of non-cellular instability below the dashed lines is subdivided into three subregions by the dotted lines. To the right of the vertical dotted line ($l>16$) below the dashed lines, we have a non-oscillatory longwave instability and to its left $(l<16)$ we have an oscillatory instability.

As is well known~\cite[pp.~477-479]{clavin2016combustion}, when $\lambda=0$ (or $\Pec=0$), the cellular instability emerges for $l<-2$ and the oscillatory instability for $l>32/3$, as can be seen in the figure. The most striking observation in the presence of Taylor dispersion ($\lambda \neq 0$) is that the cellular instability region comprises positive values of $l$ in addition to negative values. Specifically, cellular instability can now occur, according to our theoretical analysis, for any value of $l$ such that $|l|>2$. It is worth noting in particular that for $l>2$, the cellular instability can be achieved if the Peclet number is large enough. Such cellular instability appears to be more accessible in real reactive mixtures than the oscillatory instability which requires large values of $l$, namely $l>32/3$ (in the absence of heat loss). 
Another noteworthy observation pertinent to subunity Lewis number cases is that the cellular instability is hampered  by Taylor dispersion when $l<0$ and suppressed completely for sufficiently strong shear flow, $\lambda>1/2$.

\subsection{Bifurcation curves}
\label{sec:bifur}
\begin{figure}[h!]
\centering
\includegraphics[scale=0.65]{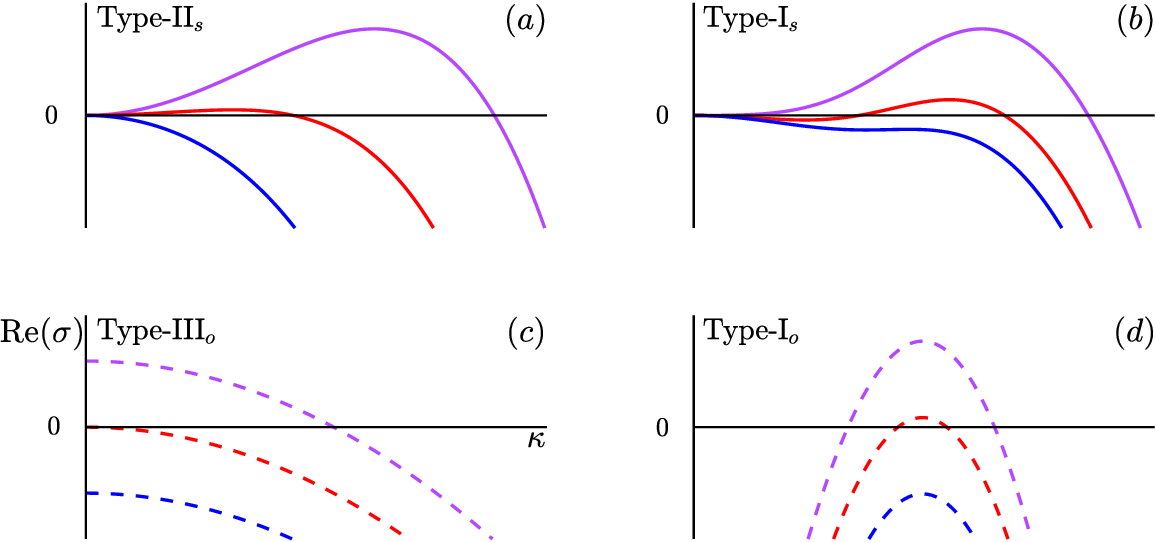}
\caption{A schematic illustration of the four types of bifurcations identified in Fig.~\ref{fig:regime}. Plotted is $\mathrm{Re}\{\sigma(\kappa)\}$ as a function of wavenumber $\kappa$ for the eigenvalue $\sigma$ which crosses during the bifurcation from the left-half to the right-half complex plane.  Solid lines indicate that $\sigma$ is real ($\mathrm{Im}(\sigma)=0$) and dashed lines indicate that $\sigma$ has a non-zero imaginary part ($\mathrm{Im}(\sigma)\neq 0$).} \label{fig:typesch}
\end{figure}

To characterize the bifurcations from stable to unstable cases in Fig.~\ref{fig:regime} occurring at the marginal condition $\mathrm{Re}\{\sigma\}=0$, we shall adopt the terminology used in~\cite[pp.75-81]{cross2009pattern} and~\cite{cross1993pattern}. According to this terminology, the types of bifurcations encountered, as denoted in  Fig.~\ref{fig:regime}, are type-I$_s$, type-II$_s$, type-I$_o$ and type-III$_o$ bifurcations. These are illustrated schematically in Fig.~\ref{fig:typesch} where the real growth rate $\mathrm{Re}\{\sigma(\kappa)\}$ is plotted for the eigenvalue $\sigma$ which crosses during the bifurcation from the left-half to the right-half complex plane. In the terminology adopted, the subscript $s$ refers to a stationary or non-oscillatory bifurcation, also known as a zero-eigenvalue bifurcation and corresponds to $\sigma=0$. Similarly, the subscript $o$ refers to an oscillatory bifurcation and corresponds to $\mathrm{Re}(\sigma)= 0$ with $\mathrm{Im}(\sigma)\neq 0$. 

The bifurcation curves in Fig.~\ref{fig:regime} are determined as follows. The bifurcation curve labelled type-II$_s$ in Fig.~\ref{fig:regime} is obtained from the condition $d^2\sigma/d\kappa^2|_{\kappa=0}$, which is clear from Fig.~\ref{fig:typesch}(a), leading to the explicit relation 
\begin{equation}
    \lambda = \frac{l+2}{2l}  \quad \text{for} \quad l\leq -2 \quad \text{and}\quad 2<l\leq 6. \label{type2}
\end{equation}
Note that this expression is applicable in the domains $l\leq -2$ and $ 2<l\leq 6$; for $l>6$, the boundary between  stability and instability regions corresponds to different types of bifurcation curves, as indicated in Fig.~\ref{fig:regime}. Further details pertaining to this type-II$_s$ bifurcation are given in $\S$\ref{sec:weak} mainly dedicated to the derivation of a Kuramoto-Sivashinsky equation.

We turn now to the type-I$_s$ bifurcation curve identified in Fig.~\ref{fig:regime}. The curve is obtained from the conditions $\sigma=0$ and $d\sigma/d\kappa=0$ as seen in Fig.~\ref{fig:typesch}(b), which determine at the bifurcation the wavenumber\footnote{An explicit relation for $\kappa_m(l)$ can be obtained by substituting the expression for $\lambda$ in~\eqref{type1} into the equation $\kappa_m^2= [(\lambda l +2 + \sqrt{(\lambda l + 2)^2-6l})/12]^2 - 1/4$, which follows from the condition $d\sigma/d\kappa=0$. From this relation, we can deduce that $\kappa_m\sim \sqrt{l-6}$ as $l\rightarrow 6^+$.} $\kappa=\kappa_m\neq 0$ and provide the relation
\begin{equation}
    \lambda = \frac{2}{l}(\sqrt{2l+4}-2) \quad \text{applicable for} \quad 6\leq l\leq 4(1+\sqrt 3). \label{type1}
\end{equation}
We note that the type-I$_s$ and type-II$_s$ boundary curves introduced meet at a \textit{tricritical point}
\begin{equation}
    (l,\lambda) = (6,2/3).
\end{equation}
This terminology is borrowed from phase transition theory~\cite[p.~493]{landau2013statistical} where a tricritical point refers to a point where first-order and second-order phase transition curves meet; such transition curves are mathematically analogous to our type-I$_s$ and type-II$_s$ bifurcation curves.\footnote{As in phase transition theory~\cite[p.~496]{landau2013statistical}, the curves of type-I$_s$~\eqref{type1} and type-II$_s$~\eqref{type2} are continuous and have continuous first derivatives but discontinuous second and higher-order derivatives at the tricritical point.} 

Furthermore, the type-III$_o$ bifurcation curve is determined from the requirements $\mathrm{Re}\{\sigma\}=0$, $d\mathrm{Re}\{\sigma\}/d\kappa=0$ at $\kappa=\kappa_m=0$; see Fig.~\ref{fig:typesch}(c). The curve is found to be given by the equation
\begin{equation}
    l = 4(1+\sqrt 3) \quad \text{applicable for} \quad \lambda \in  [0.3902,0.5646],
\end{equation}
which corresponds to a vertical line segment in Fig.~\ref{fig:regime}. The fact that this curve is vertical is straightforward consequence of the dispersion relation~\eqref{disp} being independent of $\lambda$ when $\kappa=0$. 

Finally, the type-I$_o$ bifurcation curve satisfies the conditions $\mathrm{Re}\{\sigma\}=0$, $d\mathrm{Re}\{\sigma\}/d\kappa=0$ at $\kappa=\kappa_m\neq 0$;  see Fig.~\ref{fig:typesch}(d). This curve is computed numerically and is found to extend from the point $(\lambda,l)=(0,\frac{32}{3})$ to the point $(\lambda,l)=(0.3902, 4(1+\sqrt 3))$.

\subsection{Further mathematical implications and weakly nonlinear analysis}
\label{sec:weak}
The regime diagram of Fig.~\ref{fig:regime} suggests a rich variety of mathematical behaviours in the vicinity of the bifurcation curves and in particular near the tricritical point. Indeed, in the vicinity of the bifurcation curves, weakly nonlinear analyses can be carried out accounting for the coupling between Taylor dispersion and the thermo-diffusive instabilities. For example, in the vicinity of type-II$_s$ bifurcation curve, a Kuramoto-Sivashinsky (KS) equation can be derived and, in the neighbourhood of type-I$_s$ bifurcation curve, a Swift-Hohenberg (SH) equation can be derived~\cite{cross1993pattern}. More interestingly, near the tricritical point, sixth-order (in space) partial differential equations can be obtained, describing the flame evolution in the weakly nonlinear regime. These mathematical aspects are discussed elsewhere~\cite{rajamanickam2023tricritical}. Here we shall only address the flame dynamics near the type-II$_s$ bifurcation curve, which is classically described by a KS equation. The KS equation in our case takes the form
\begin{equation}
    f_t + \frac{p^2-1}{2} (l-l_c) f_{xx} + \frac{2(p^2+1)^2(p^2-2)}{p^2-1} f_{xxxx} + \frac{p^2+1}{2}f_x^2=0 \label{KS}
\end{equation}
where 
\begin{equation}
    l_c = \frac{2(p^2+1)}{p^2-1} = \frac{2}{2\lambda-1}. \label{lc}
\end{equation}

Note that the nonlinear term in~\eqref{KS} may be derived using a semi-heuristic kinematic argument as explained in~\cite{sivashinsky1983instabilities}, and as done in~\cite{daou2023flame} in the similar problem of flames propagating in a longitudinal direction. The details of the derivation will not be repeated here as they are similar to those in~\cite{daou2023flame}.

As for the linear part of~\eqref{KS}, this can be simply obtained from the dispersion relation~\eqref{disp}. To this end, we note that equation~\eqref{disp} has always a real root $\sigma(k)$ such that $\sigma(0)=0$ and that the  Taylor expansion of $\sigma(k)$ for small values of $k$ is given by 
\begin{equation}
    \sigma = \alpha_2 k^2- \alpha_4 k^4  + \cdots \label{taylor}
\end{equation}
where $\alpha_2$ and $\alpha_4$ are given by
\[
\alpha_2 = \frac{p^2-1}{2} (l-l_c) \quad \text{and} \quad \alpha_4 = \frac{2(p^2+1)^2(p^2-2)}{p^2-1}.
\]
These are the coefficients of $f_{xx}$ and $f_{xxxx}$ in~\eqref{KS}, since the linear part of~\eqref{KS} is equivalent to~\eqref{taylor} in the case of normal modes $f(x,t) \propto e^{\sigma t+ ikx}$.

It is worth noting that the onset of the (type-II$_s$) instability corresponds to the condition $\alpha_2=0$ which is equivalent to the condition $d^2\sigma/dk^2|_{k=0}=0$ used earlier in $\S$\ref{sec:bifur}. This is so, provided the coefficient $\alpha_4$ of the fourth-derivative term is positive. Indeed, if $\alpha_4$ were negative, then the critical wavenumber $k_c$ at the onset of instability would be necessarily non-zero, which contradicts the requirement $k_c=0$ of a type-II$_s$ bifurcation. The condition $\alpha_4>0$ requires $p>\sqrt 2$ or $p<1$, that is equivalently, $\lambda>2/3$ or $\lambda<1/2$. Therefore the critical condition given by~\eqref{lc} is only applicable in this range of $\lambda$ (or $p$), which determines the black type-II$_s$ bifurcation curve in Fig.~\ref{fig:regime}. For $\lambda<2/3$ and $l>0$, cellular flames still exist but the transition from the stable to the unstable region occurs through a type-I$_s$ bifurcation, represented by the red curve in~Fig.~\ref{fig:regime}.

\section{Computational results for finite Zeldovich number $\beta$}
\label{sec:numerical}

The theoretical results discussed so far are all based on the dispersion relation~\eqref{disp} obtained in the asymptotic limit $\beta\to \infty$. In this section, we shall carry out computations with a finite value of $\beta$ in order to assess the applicability of theoretical predictions, at least qualitatively, and also to examine the nonlinear evolution of unstable flames. To this end, we consider the problem consisting of  governing equations~\eqref{yF} and~\eqref{theta} and boundary conditions~\eqref{BCunburnt}-\eqref{BCburnt}, adopting the numerical values $\beta=10$ and $\alpha=0.85$, while varying the parameters $\Lew$ and $p^2$ (or equivalently $\lambda=p^2/(1+p^2)$).

 \begin{figure}[h!]
\centering
\includegraphics[scale=0.6]{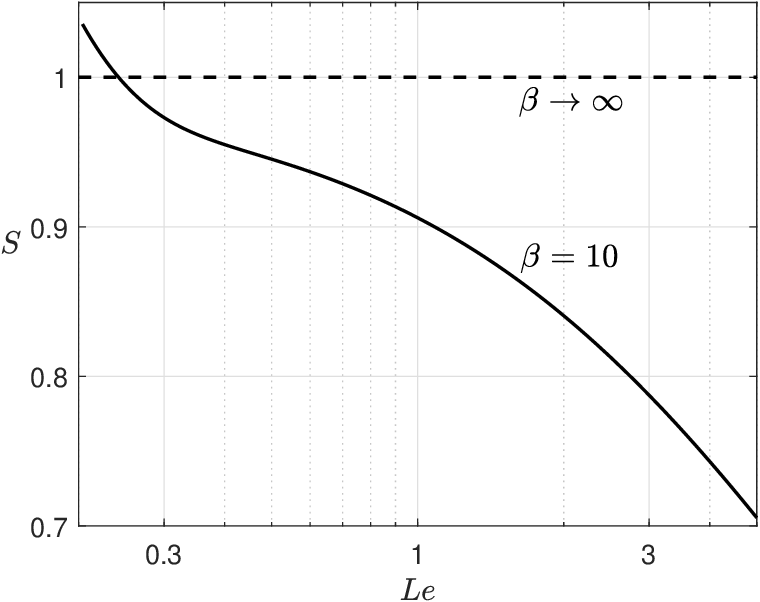}
\caption{The scaled planar-flame burning speed $S=S_L/S_L^0$  versus the Lewis number $\Lew$, for $\beta=10$ and $\alpha=0.85$.} 
\label{fig:finitebetaS}
\end{figure}
\begin{figure}[h!]
\centering
\includegraphics[scale=0.8]{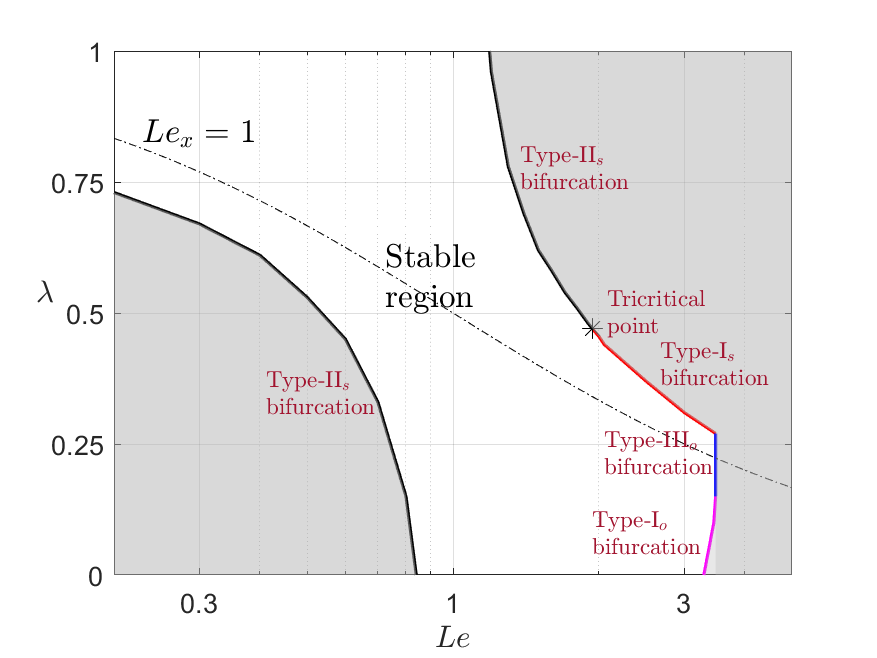}
\caption{Stability regime diagram of a premixed flame propagating transversely to a shear flow in the $\Lew$-$\lambda$ plane. The figure is based on the computation of the eigenvalues of problem~\eqref{yfeig}-\eqref{eigbc} with $\beta=10$ and $\alpha=0.85$. The grey region corresponds to unstable flames and the white to stable flames. Compare with Fig. \ref{fig:regime}, where analogous notations are used.} 
\label{fig:finitebetastability}
\end{figure}

\subsection{Preliminary considerations}
\label{sec:prelim}
Before presenting the finite-$\beta$ numerical results in the next two subsections, we make a few preliminary remarks which will facilitate the discussion and interpretation of the results. We begin by noting upon examining equations~\eqref{yF} and~\eqref{theta} that the presence of a flow-dependent \textit{effective Lewis number in the $x$-direction}, as noted in~\cite{daou2018taylor,rajamanickam2022thick}, namely,
\begin{equation}
     \Lew_{x}=\frac{\Lew(1+\gamma \Pec^2)}{1+\gamma \Pec^2\Lew^2} = \frac{\Lew(1+p^2)}{1+p^2\Lew^2} = \frac{\Lew}{1+\lambda(\Lew^2-1)},  \label{Lex}
\end{equation}
where use has been made of the definitions $p^2=\gamma\Pec^2$ and $\lambda=p^2/(1+p^2)$. 

An important implication of formula~\eqref{Lex} is that $\Lew_x\rightarrow \Lew$ as $\Pec\rightarrow 0$ and $\Lew \rightarrow 1/\Lew$ as $\Pec\rightarrow \infty$. In other words, the effect of a shear flow is such that a weakly diffusing reactant ($\Lew>1$) appears effectively as strongly diffusing (in the $x$-direction, $\Lew_x<1$) and a strongly diffusing reactant ($\Lew<1$) appears effectively as weakly diffusing ($\Lew_x>1$) provided the Peclet number is large enough, more precisely when $\Pec$ exceeds the value $1/\sqrt{\gamma\Lew}$. In fact, this value is determined from the fact that $\Lew_x=1$ irrespective of $\Lew$ when $\Pec=1/\sqrt{\gamma\Lew}$, which follows from \eqref{Lex}.

\subsection{Linear stability analysis based on eigen-boundary value problem}
We begin by examining the linear stability of the planar flame solution with the finite value $\beta=10$ adopted. 
This base solution satisfies equations~\eqref{yF}-\eqref{theta} with $\pfi{}{t}=\pfi{}{x}=0$. The scaled burning speed $S$, which is equal to unity in the limit $\beta\rightarrow \infty$, is now computed numerically and plotted as a function of the Lewis number $\Lew$ in Fig.~\ref{fig:finitebetaS}. If we denote the dependent variables of the base state as $\bar y_F(y)$ and $\bar\theta(y)$, then the stability of the base solution to small perturbations can be studied by introducing 
\begin{equation}
    \begin{bmatrix}
 y_F\\
\theta
\end{bmatrix}=\begin{bmatrix}
\bar y_F\\
\bar\theta
\end{bmatrix}
+  e^{ik\sqrt{1+p^2}x+\sigma t} \begin{bmatrix}
\tilde y_F\\
\tilde\theta
\end{bmatrix}
\end{equation}
into the problem~\eqref{yF}-\eqref{BCburnt}. The linear stability is then described by the eigen-boundary value problem 
\begin{align}
     \frac{1}{\Lew} \frac{d^2\tilde y_F}{dy^2} - S \frac{d\tilde y_F}{dy} - \frac{k^2}{\Lew_x} \tilde y_F -  \left\{\tilde y_F + \frac{\beta \bar y_F \tilde \theta}{[1-\alpha(1-\bar\theta)]^2}\right\} \frac{\beta^2}{2\Lew}  \exp\left[\frac{\beta(\bar\theta-1)}{1+\alpha(\bar\theta-1)}\right]   = \sigma\tilde{y}_F,\label{yfeig}\\
   \frac{d^2\tilde \theta}{dy^2} - S \frac{d\tilde \theta}{dy} - k^2 \tilde \theta +  \left\{\tilde y_F + \frac{\beta \bar y_F \tilde \theta}{[1-\alpha(1-\bar\theta)]^2}\right\} \frac{\beta^2}{2\Lew}  \exp\left[\frac{\beta(\bar\theta-1)}{1+\alpha(\bar\theta-1)}\right]   = \sigma\tilde\theta  
\end{align}
with 
\begin{align}
    \tilde y_F=0, \quad \tilde\theta=0 &\quad \text{as} \quad y\rightarrow -\infty \\
    \tilde y_F=0, \quad \frac{d\tilde \theta}{dy}=0 &\quad \text{as} \quad y\rightarrow +\infty. \label{eigbc}
\end{align}
It is worth noting the presence in the equations of the longitudinal effective Lewis number $\Lew_x$ defined in~\eqref{Lex}, in addition to the Lewis number $\Lew$. The problem possesses a discrete spectrum of eigenvalues $\sigma$, in which the eigenvalue with the maximum real part (growth rate) determines the stability of the base solution. The corresponding computational results  are summarized in Fig.~\ref{fig:finitebetastability}, where the stability-instability regions are delimited in the $\Lew$-$\lambda$ plane. The solid lines in the figure separates the stable (white) region from the unstable (grey) regions. The dash-dotted line represents the condition $\Lew_x=1$, which  takes the form $\lambda=1/(\Lew+1)$ on using $\lambda=p^2/(1+p^2)$ in~\eqref{Lex}.

\begin{figure}[h!]
\centering
\advance\leftskip-1.6cm
\includegraphics[scale=0.6]{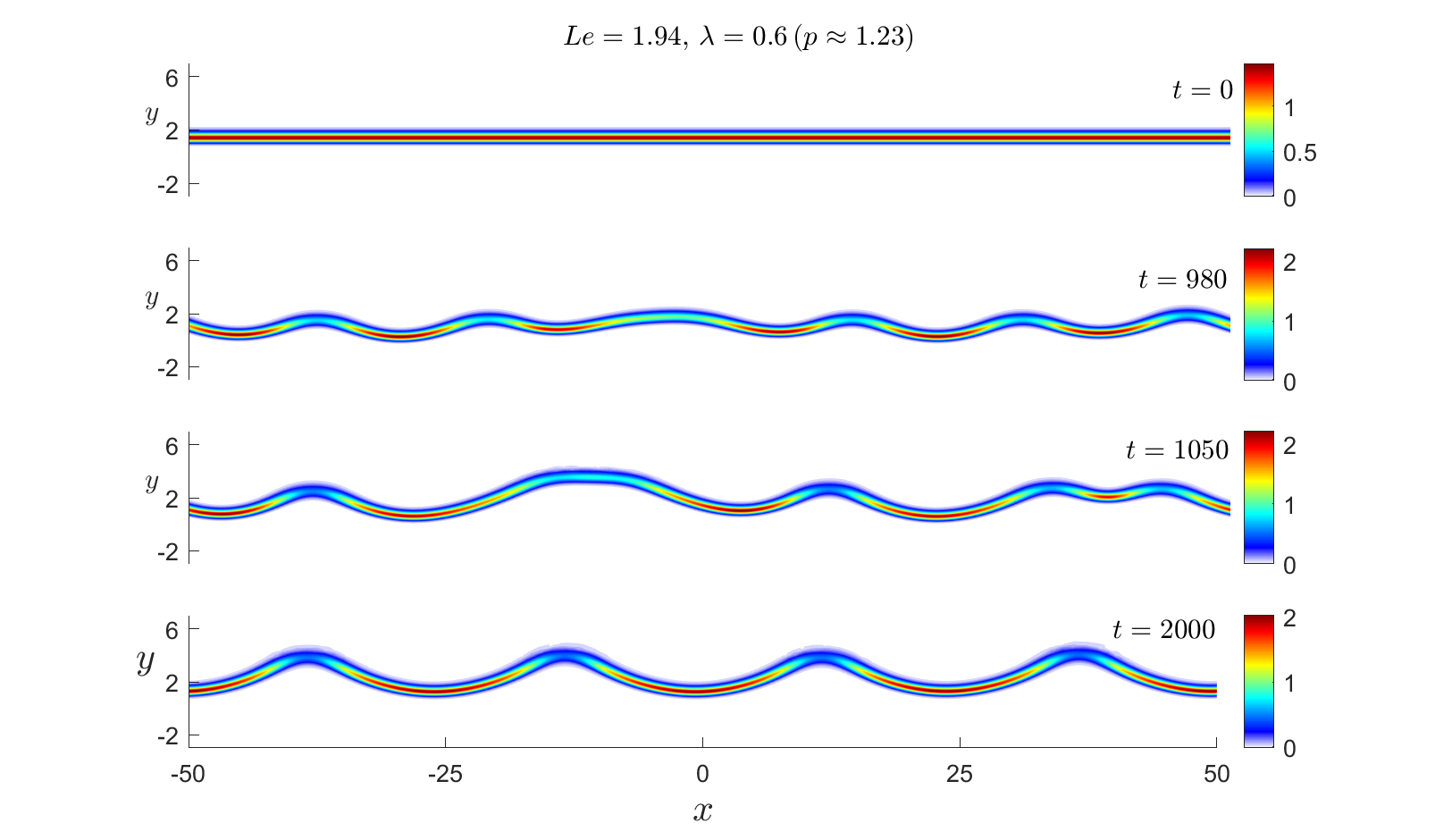}
\caption{Fields of reaction rate $\omega$ at selected values of time $t$, computed for $(\Lew,\lambda)=(1.94,0.6)$, $\beta=10$ and $\alpha=0.85$. The initial condition for the time-dependent calculations corresponds to a steady, planar premixed flame, computed numerically.} 
\label{fig:reactrate1}
\end{figure}
\begin{figure}[h!]
\centering
\advance\leftskip-1.6cm
\includegraphics[scale=0.6]{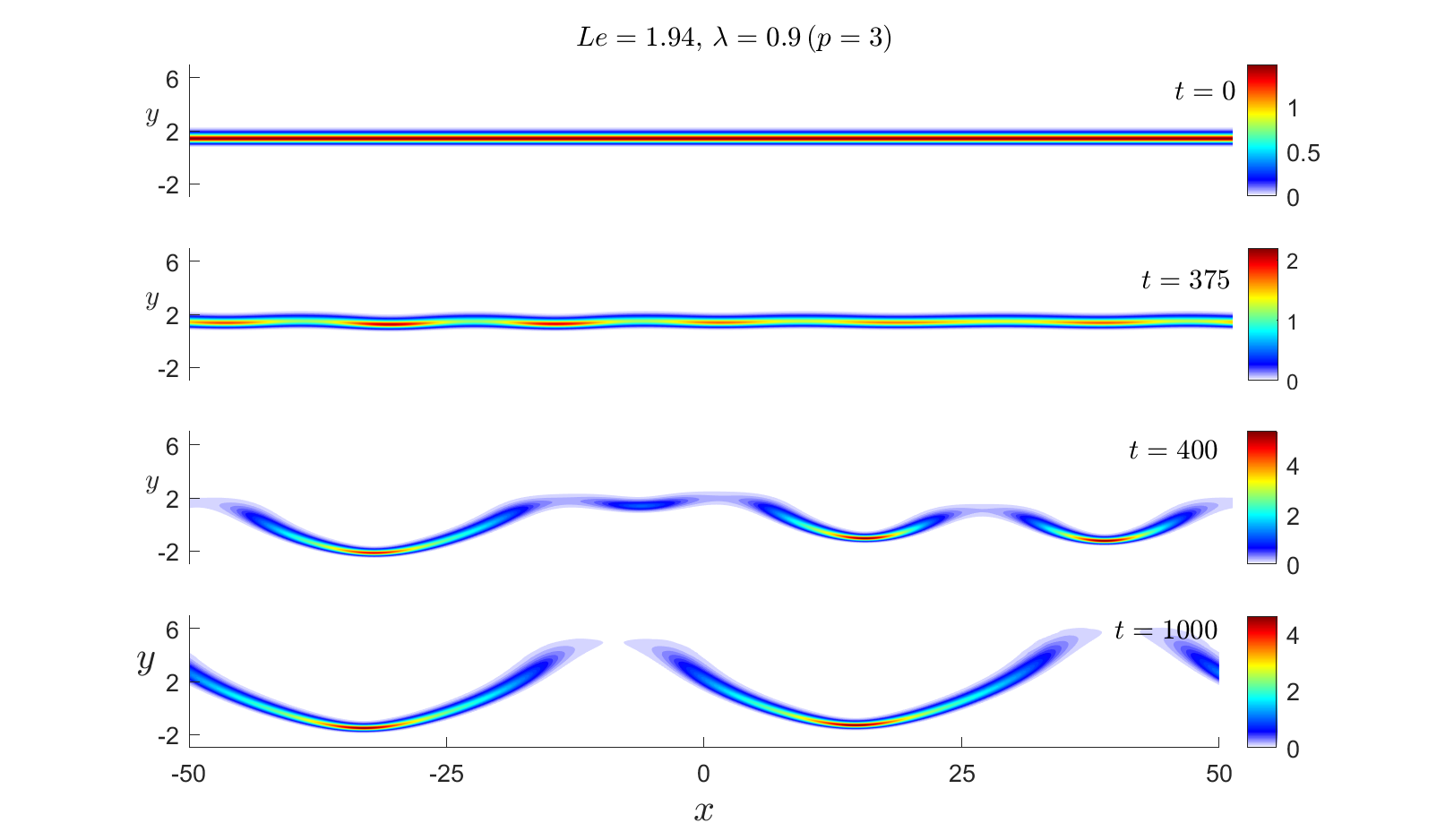}
\caption{Fields of reaction rate $\omega$ at selected values of time $t$, computed for $(\Lew,\lambda)=(1.94,0.9)$, $\beta=10$ and $\alpha=0.85$. The initial condition for the time-dependent calculations corresponds to a steady, planar premixed flame, computed numerically.} 
\label{fig:reactrate2}
\end{figure}

Comparing Fig.~\ref{fig:finitebetastability}, computed for $\beta=10$, with Fig.~\ref{fig:regime} corresponding to $\beta\rightarrow \infty$, the following conclusions can be drawn. First, we note the good qualitative agreement between the two figures, notably, regarding the bifurcation curves separating the stability from instability regions. In particular note the presence of a tricritical point at $(\Lew,\lambda)\approx (1.94,0.47)$, identified in Fig.~\ref{fig:regime}.  Second, the computations confirm the existence of the cellular instability region for $\Lew>1$, an original finding of the asymptotic analysis.  More precisely, two cellular instability regions appear in Fig.~\ref{fig:finitebetastability}, whose locations are compatible with the conditions $\Pec <1/\sqrt{\gamma\Lew}$ when $\Lew<1$ and $\Pec >1/\sqrt{\gamma\Lew}$ when $\Lew>1$, or equivalently $\Lew_x<1$ for any $\Lew$, as argued in $\S$\ref{sec:prelim}. Further the flame is found to be stable for all values of $\lambda$ when $\Lew\in(0.84,1.19)$ in the case $\beta=10$, while the equivalent stability range predicted by the asymptotic analysis, $l\in(-2,2)$, provides the stability range $\Lew=1+l/\beta\in(0.8,1.2)$ for $\beta=10$.

\subsection{Time-dependent numerical simulations}
Finally, we present time-dependent numerical simulations simply to illustrate the occurrence of the cellular instability for $\Lew>1$ and the flame long-time evolution. The simulations are based on the numerical solution of the time-dependent problem~\eqref{yF}-\eqref{BCburnt}, with periodic conditions in the $x$-direction, starting from an initial condition corresponding to a steady, planar premixed flame. The computations are carried out using COMSOL Multiphysics software as described in~\cite{daou2023flame,rajamanickam2022thick}. Two cases are considered corresponding to $(\Lew,\lambda)=(1.94,0.6)$ and $(\Lew,\lambda)=(1.94,0.9)$, whose results are shown in Fig.~\ref{fig:reactrate1} and Fig.~\ref{fig:reactrate2}, respectively. Shown are reaction-rate $(\omega)$ fields at selected values of time $t$. These illustrate the development of the instability and the flame evolution into a cellular structure, which settles into an apparently stable state at large times.

\section{Conclusions}
\label{sec:conclusions}

In this paper, we have examined the effect of Taylor dispersion, or shear-enhanced diffusion on the stability of a premixed flame, propagating in a direction transverse to that of unidirectional shear flow. A simple thermo-diffusive model is adopted corresponding to a Hele-Shaw channel, whose walls are assumed to be adiabatic and closely spaced. In this configuration, a simple shear flow is prescribed, corresponding for example to a Couette flow. Upon depth-averaging, the problem is governed by two-dimensional  transport equations in which diffusion is anisotropic with shear-enhanced diffusion coefficients in the longitudinal flow direction. A linear stability analysis is carried out in the limit $\beta\to\infty$, where $\beta$ is the Zeldovich number. A simple dispersion relation~\eqref{disp} is derived analytically involving the Lewis number $\Lew$ and the Peclet number $\Pec$ as parameters.
A stability regime diagram (Fig.~\eqref{fig:regime}) is constructed in  the $\Lew$-$\Pec$ plane, which categorises the various instabilities and bifurcations encountered. 

A remarkable finding is the demonstration that the classical cellular instability, which is usually expected to occur in $\Lew<1$ mixtures, exists now only for $\Lew>1$ mixtures when the Peclet number exceeds a critical value. \blue{The necessary conditions needed to observe cellular flames in $\Lew>1$ mixtures include  that the geometry must be slender such as in the case of a Hele-Shaw channel, that the Peclet number $\Pec$ must exceed a critical value above  $1/\sqrt{\gamma \Lew}$ and that the flame must be aligned with the flow direction.} Furthermore, the cellular instability identified for $\Lew>1$ is shown to occur either through a finite-wavelength stationary bifurcation (also known as type-I$_s$) or through a longwave stationary bifurcation (also known as type-II$_s$). In the weakly nonlinear regime, a Kuramoto-Sivashinsky equation in the vicinity of type-II$_s$ bifurcation is derived. Moreover, it is found that the oscillatory instability, usually encountered in $\Lew>1$ mixtures persists under the influence of Taylor dispersion if the Peclet number is below a critical value and disappears above this value.

The stability results aforementioned, which follow from the dispersion relation obtained in the limit $\beta\to\infty$, are complemented by numerical computations carried out for a finite value of the Zeldovich number. The computations involve the determination of the eigenvalues of a linear stability boundary-value problem and numerical simulations of the time-dependent governing partial differential equations. The numerical results are found to be in good qualitative agreement with the analytical predictions. In particular, a stability regime diagram (Fig.~\eqref{fig:finitebetastability}) computed for $\beta=10$ is found to be consistent with the stability regime diagram of Fig.~\eqref{fig:regime}, corresponding to $\beta\to\infty$.

To close this paper, we note that the stability results obtained for the case of transverse flame propagation differ markedly from those obtained for flames propagating in the longitudinal direction~\cite{daou2021effect}. For example, the appearance of cellular instability in $\Lew>1$ mixtures when the Peclet number is above a critical value, does not occur in the case of longitudinal propagation. In fact, in the latter case, only stable flames are encountered when $\Lew>1$ and $\Pec\gg 1$. A natural follow-up of the current work is to extend the stability analysis to flames propagating in an arbitrary direction relative to that of the shear flow. \blue{It is also worthwhile to investigate the influence of heat losses on our predictions, following~\cite{daou2023flame}, since such influence is significant in practice.} Interestingly, the cellular instability identified herein for $\Lew>1$  mixtures  has been recently shown to also take place for diffusion flames aligned with the direction of the shear flow~\cite{rajamanickam2023stability}. This unexpected instability has been proposed in~\cite{rajamanickam2023stability} as a plausible mechanism for the formation of \textit{diffusion flame streets} observed in experiments~\cite{miesse2005diffusion,miesse2005experimental,xu2009studies}. It is  
desirable to have similar experiments in the premixed case to test the presence of the cellular instability identified herein for $\Lew>1$ mixtures in the presence of a strong shear flow.

\section*{Acknowledgements}
This research was supported by the UK EPSRC through grant EP/V004840/1.

\bibliographystyle{tfq}
\bibliography{interacttfqsample}

\end{document}